# Resistivity of Graphene Nanoribbon (GNR) Interconnects

Raghunath Murali[1], Kevin Brenner, Yinxiao Yang, Thomas Beck, and James D. Meindl

Nanotechnology Research Center, Georgia Institute of Technology, Atlanta GA

*Abstract*—**Graphene nanoribbon interconnects are fabricated, and the extracted resistivity is compared to that of Cu. It is found that the average resistivity at a given line-width (18nm<W<52nm) is about 3X that of a Cu wire, whereas the best GNR has a resistivity** *comparable* **to that of Cu. The conductivity is found to be limited by impurity scattering as well as LER scattering; as a result, the best reported GNR resistivity is 3X the limit imposed by substrate phonon scattering. This study reveals that even moderate-quality graphene nanowires have the potential to outperform Cu for use as on-chip interconnects.**

## I INTRODUCTION

GRAPHENE as an electronic material has been receiving much attention as a possible replacement for Silicon CMOS technology. In addition to its use as a switching device, Graphene can also be used as an interconnect material – a truly monolithic system can be constructed using graphene for both transistors and interconnects [1]. Compared to Silicon and even III-V semiconductors, Graphene has superior mobility. Ballistic transport in graphene makes it attractive not only for use as transistors but also for interconnects.

Graphene was first isolated in 2004 and since then many properties have been confirmed experimentally including high mobility, ballistic transport, linear E-k relationship, and a width-dependent transport gap. Transistors fabricated from graphene nanowires have shown impressive on-off ratio [2], [3]. For interconnect applications, graphene has shown interesting properties in terms of its temperature coefficient [4], and use as a thermal interface material [5]. Theoretical projections for GNRs for use as interconnects have been made in [6] and GNRs are predicted to out-perform Cu for interconnect applications. In [6], single-layer GNRs are shown to result in a lower resistance per unit length than 1:1 AR Cu, for line-width (W) less than 8 nm. There has been little experimental work on the electrical characterization of graphene for use as an interconnect material. While there is some experimental data for wide graphene ribbons [7], there has not been a methodical investigation of resistivity for narrow width GNRs (W<100 nm). This work will focus on characterizing graphene resistivity for narrow-width GNRs, and will compare the resistivity of GNRs to that of Cu.

---

[1] Email: raghu@gatech.edu

## II EXPERIMENT

Few layer graphene is used as the starting material. Graphene layers are flaked from large graphite pieces (Kish Graphite, Toshiba Ceramic Co.) using a Scotch tape, and adhered onto an oxidized Si substrate with an oxide thickness of 300 nm. The Si substrate is degenerately doped for use as a back-gate. Using electron beam lithography, contact pads are formed by a metal lift-off process using Ti/Au (10/90nm) as the contact metal. This is followed by characterization of the 2D graphene for contact resistance, Dirac point, and carrier modulation. A second layer of lithography defines nanometer-wide channels with dimensions in the range 18nm<W<52nm, and 0.2µm<L<1µm. A low-power oxygen plasma etch is used to transfer resist patterns into the channel.

Electrical measurements were performed with standard lock-in techniques; excitation currents of 5nA-100nA were used in four-point probe measurements to extract contact resistance. A HP 4156 semiconductor parameter analyzer was used to perform low-bias measurements, along with a back-gate sweep. Tests for ohmic contacts were performed at voltages down to a few micro-volts, and there was no indication of a Schottky barrier. High-resolution SEM imaging revealed the dimensions of patterned GNRs. Monolayer and bilayer graphene are identified by confocal micro-Raman imaging. For three or more layers, AFM scanning is used to estimate the number of graphene layers. The optical visibility of few-layer graphene is a quick technique in differentiating between monolayer and few-layer graphene. Fig. 1. shows a representative device; in this device, there are 10 GNRs between each pair of electrodes.

Contact resistance was extracted for each device at various stages of processing. The contact resistance was found to be 30 µΩ-cm$^2$ for most devices, and did not change after the second lithography step or after the plasma etch. Back-gated measurements of conductance modulation reveals a small negative Dirac point shift after HSQ spin and e-beam lithography, and a large positive Dirac point shift after plasma etch. Unlike true metals, a semi-metal such as graphene shows significant variation in conductivity with application of a back-gate bias (since this causes a shift in the Fermi level). Thus, it is important to measure resistance at the same electron density across different GNRs to ensure a fair comparison. This is done by making $V_g$-$V_D$ ($V_g$ is the gate-voltage, and $V_D$ is the Dirac voltage) the same across GNRs when extracting device parameters. The back-gate capcitance is 11.5 nF/cm$^2$, and for an electron density of n=5x10$^{12}$ cm$^{-2}$, this translates to a $V_g$-$V_D$ of 70V. All resistance and resistivity measurements in this work are thus measured at $V_g$-$V_D$ =-70V (hole carriers) so that the corresponding carrier density makes the GNR operate in the metallic regime. This translates to a Fermi energy of 63 meV [8] for 2D graphene.

A total of 18 devices, each with 10 parallel GNRs, were measured; these devices were selected from a larger set of samples based on their resistivity. For ease of comparison to Cu, 3D resistivity ($ρ_{3d}$) is calculated for these devices; it is more common to calculate the 2D resistivity for a 2D material like graphene. Fig. 2 shows the resistivity of the various devices as a function of GNR width. Most of the data is clustered between 15-25 µΩ-cm. Also shown for comparison is the Cu resistivity as projected by ITRS 2007 [9]. Note that little experimental data exists for narrow Cu lines and ITRS 2007 projections are based on extending current model parameters to narrow line-widths. Line-edge roughness (LER) and liner scaling becomes increasingly challenging for W<30 nm, and will lead to additional increases in effective Cu resistivity; thus, ITRS 2007 projections for Cu resistivity are optimistic but are used nonetheless since they can be thought of as the best-case scenario for a Cu line.

## III ANALYSIS

There have been a few published results on 2D resistivity for wide and narrow GNRs; it is informative to compare relevant published data from [2, 3, 10-12] against the current results, Fig. 2. All measurements in this work were done at 300 K; some of the data points from previously published data were extracted from low-temperature measurements but nevertheless provide a useful comparison. The resistivity from [2] is between 16-30 µΩ-cm, though the measurements were made at 200K. For W<100 nm, resistivity from other previously published data is more spread out. Resistivities of GNRs from this study are some of the lowest values reported for narrow GNRs. But the GNRs are still 2-3X less conductive than Cu wires.

The 18 devices shown in Fig. 2 have a set of ten GNRs in parallel. Critical dimension (CD) uniformity, LER, and the starting graphene material would cause individual GNRs to have different properties compared to one another. The ribbon-to-ribbon non-uniformity is masked somewhat since ten GNRs are measured in parallel. To extract properties of single GNRs (rather than a parallel set), a large number of GNRs would have to be fabricated and their properties extracted to obtain a statistically significant set of resistivity data. Because of the finite size of graphene flakes (usually less than 20 µm$^2$), it is not possible to fabricate a large number of GNRs (that can be probed one at a time) on the same flake. By employing techniques applied to CNTs [13], it is possible to use the device shown in Fig. 1 to extract the performance of individual GNRs.

A HP4156 semiconductor analyzer is used to apply a voltage ramp between two electrodes with 10 GNRs in parallel. Due to increasing current density in the GNRs, there is a voltage at which a GNR breaks down, resulting in a visible drop in current. The device testing is stopped at this point, and the voltage ramp is repeated from 0V. Successive GNR breakdowns occur at around the same voltage as for the first breakdown event. By recording the difference in conductance between two successive breakdown events, the individual GNR conductance can be extracted. It is also found that if the voltage-ramp steps are small enough (2 mV), it is possible to avoid multiple GNR breakdowns in a single event. The contact resistance does not change after each event, and this indicates the robust nature of the contact metallization. Back-gated conductance modulation is extracted for each GNR, and the modulation does not significantly change from one GNR to another – this means that the GNRs are of similar metallicity. All GNRs studied in this work showed an impressive breakdown current density of 5-20 $\times 10^8$ A/cm$^2$ pointing to the superior electromigration performance of GNRs. A more detailed discussion of breakdown current density and its correlation with resistivity for various GNR dimensions will be given elsewhere.

It is found that there is a significant difference in resistance from one GNR to another, even on the same flake. Fig. 3 plots the range of GNR resistivity extracted for four different devices, each with a different line-width. For each width, the best, worst and mean values of resistivity are shown. The best GNR has a resistivity comparable to that of Cu. For monolayer 2-D graphene on SiO$_2$, phonon scattering limits room temperature resistivity to about 1.2 μΩ-cm (at n=5x10$^{12}$ cm$^{-2}$). Thus, the best GNR is 3X less conductive than this limit (1.2 μΩ-cm).

Graphene on SiO$_2$ has various scattering mechanisms limiting its conductivity (1) intrinsic scattering, which limits mobility to 200,000 cm$^2$/V-s and is seen in suspended graphene [14], (2) extrinsic scattering due to SiO$_2$ phonons, which imposes a carrier mobility limit of 40,000 cm$^2$/V-s at n=1x10$^{12}$ cm$^{-2}$ and T=300K [15], (3) impurity scattering, and (4) LER scattering. It is possible to estimate the contribution of impurity scattering in GNRs using the scattering theory presented in [7]. The impurity density is estimated to be $n_i$=2-19x10$^{11}$ cm$^{-2}$ for the set of devices shown in Fig. 2. This translates to an impurity limited mobility of 2,500-19,000 cm$^2$/V-s. It is possible to estimate the LER limited mobility by extracting the difference in mobility before and after plasma etch (which converts 2D graphene flakes into GNRs) using Matthiessen's rule. For the W=22 nm GNRs shown in Fig. 3, the LER limited mobility is in the range of 6000-9000 cm$^2$/V-s, and the effective GNR mobility is in the range of 4000-8000 cm$^2$/V-s; thus, GNR mobility at W=22nm could either be limited by impurity scattering or LER scattering.

In the discussion of GNR resistivity above, it has been implicitly assumed that multi-layered graphene will be readily available to fabricate GNRs – if only single or few-layer graphene is available, then a more apt comparison parameter would be resistance per unit length. Recent experiments with graphene grown on SiC substrates have shown that truly non-interacting multi-layer graphene films of tens of layers can be formed [16] – rotational stacking preserves the ballistic nature of carriers, and would be valuable for interconnect applications.

## IV CONCLUSION

Graphene interconnects of narrow widths have been fabricated and compared to Cu interconnects in terms of their 3D resistivity. The average GNR resistivity is higher than the projected Cu resistivity for 18nm<W<52nm. Resistivity of individual GNRs have been extracted from sets of parallel GNRs, and it is found that the best GNR (for a given width) has a resistivity comparable to a Cu-wire of the same width. An analysis of scattering mechanisms reveals that narrower GNRs are limited either by LER or impurity scattering. This work gives the first experimental evidence of the potential of narrow GNRs for use as on-chip interconnects.

## ACKNOWLEDGEMENTS

This work was supported in part by the Interconnect Focus Center (IFC) - a DARPA/SRC Focus Center, and by the Nanoelectronics Research Initiative (NRI) through the INDEX Center. The authors thank A. Neemi for helpful discussions.

REFERENCES

[1] C. Berger, Z. M. Song, X. B. Li, X. S. Wu, N. Brown, C. Naud, D. Mayou, T. B. Li, J. Hass, A. N. Marchenkov, E. H. Conrad, P. N. First, and W. A. de Heer, "Electronic confinement and coherence in patterned epitaxial graphene," *Science,* vol. 312, pp. 1191-1196, May 2006.

[2] M. Y. Han, B. Ozyilmaz, Y. B. Zhang, and P. Kim, "Energy band-gap engineering of graphene nanoribbons," *Physical Review Letters,* vol. 98, p. 206805, May 2007.

[3] X. R. Wang, Y. J. Ouyang, X. L. Li, H. L. Wang, J. Guo, and H. J. Dai, "Room-temperature all-semiconducting sub-10-nm graphene nanoribbon field-effect transistors," *Physical Review Letters,* vol. 100, p. 206803, May 2008.

[4] Q. Shao, G. Liu, D. Teweldebrhan, and A. A. Balandin, "High-temperature quenching of electrical resistance in graphene interconnects," *Applied Physics Letters,* vol. 92, p. 202108, May 2008.

[5] S. Ghosh, I. Calizo, D. Teweldebrhan, E. P. Pokatilov, D. L. Nika, A. A. Balandin, W. Bao, F. Miao, and C. N. Lau, "Extremely high thermal conductivity of graphene: Prospects for thermal management applications in nanoelectronic circuits," *Applied Physics Letters,* vol. 92, p. 151911, Apr 2008.

[6] A. Naeemi and J. D. Meindl, "Conductance modeling for graphene nanoribbon (GNR) interconnects," *IEEE Electron Device Letters,* vol. 28, pp. 428-431, May 2007.

[7] Y. W. Tan, Y. Zhang, K. Bolotin, Y. Zhao, S. Adam, E. H. Hwang, S. Das Sarma, H. L. Stormer, and P. Kim, "Measurement of scattering rate and minimum conductivity in graphene," *Physical Review Letters,* vol. 99, p. 246803, Dec 2007.

[8] F. Munoz-Rojas, J. Fernandez-Rossier, L. Brey, and J. J. Palacios, "Performance limits of graphene-ribbon field-effect transistors," *Physical Review B,* vol. 77, p. 045301, Jan 2008.

[9] Semiconductor Industry Association, "International Technology Roadmap for Semiconductors," 2007.

[10] M. C. Lemme, T. J. Echtermeyer, M. Baus, and H. Kurz, "A graphene field-effect device," *Ieee Electron Device Letters,* vol. 28, pp. 282-284, Apr 2007.

[11] Z. H. Chen, Y. M. Lin, M. J. Rooks, and P. Avouris, "Graphene nano-ribbon electronics," *Physica E-Low-Dimensional Systems & Nanostructures,* vol. 40, pp. 228-232, Dec 2007.

[12] B. Huard, N. Stander, J. A. Sulpizio, and D. Goldhaber-Gordon, "Evidence of the role of contacts on the observed electron-hole asymmetry in graphene," *Physical Review B,* vol. 78, p. 121402, Sep 2008.

[13] P. C. Collins, M. S. Arnold, and P. Avouris, "Engineering carbon nanotubes and nanotube circuits using electrical breakdown," *Science,* vol. 292, pp. 706-709, Apr 2001.

[14] K. I. Bolotin, K. J. Sikes, Z. Jiang, M. Klima, G. Fudenberg, J. Hone, P. Kim, and H. L. Stormer, "Ultrahigh electron mobility in suspended graphene," *Solid State Communications,* vol. 146, pp. 351-355, June 2008.

[15] J. H. Chen, C. Jang, S. D. Xiao, M. Ishigami, and M. S. Fuhrer, "Intrinsic and extrinsic performance limits of graphene devices on SiO2," *Nature Nanotechnology,* vol. 3, pp. 206-209, Apr 2008.

[16] J. Hass, F. Varchon, J. E. Millan-Otoya, M. Sprinkle, N. Sharma, W. A. De Heer, C. Berger, P. N. First, L. Magaud, and E. H. Conrad, "Why multilayer graphene on 4H-SiC(000(1)over-bar) behaves like a single sheet of graphene," *Physical Review Letters,* vol. 100, p. 125504, Mar 2008.

FIGURES

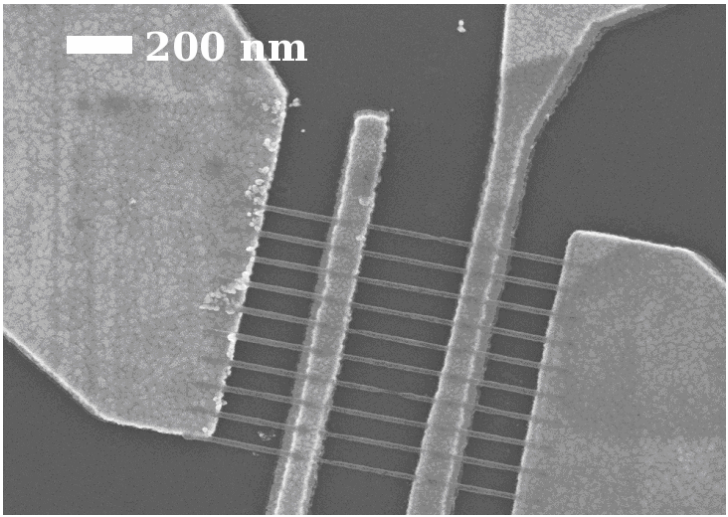

Figure 1: SEM image of a set of 10 GNRs between each electrode pair. The GNRs (below HSQ lines) are 22 nm wide between the middle electrode pair.

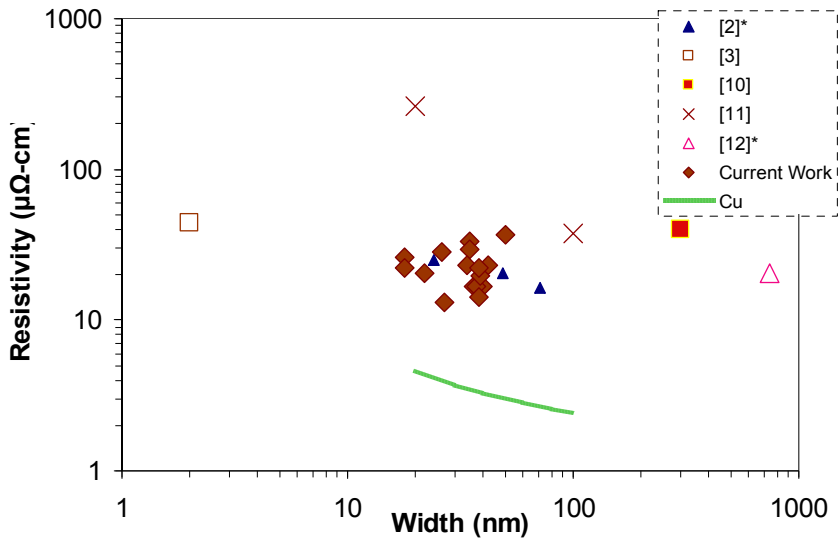

Figure 2: Resistivity of narrow-width GNRs: resistivity of eighteen devices (each with ten GNRs in parallel) from the present study are shown. Also shown is previously published data from [2, 3, 10-12]. The Cu resistivity data is from [9]. A "*" in the legend indicates that the corresponding measurement was made at below room-temperature. The data from [2] is extracted at 200K, and data from [12] at 4K.

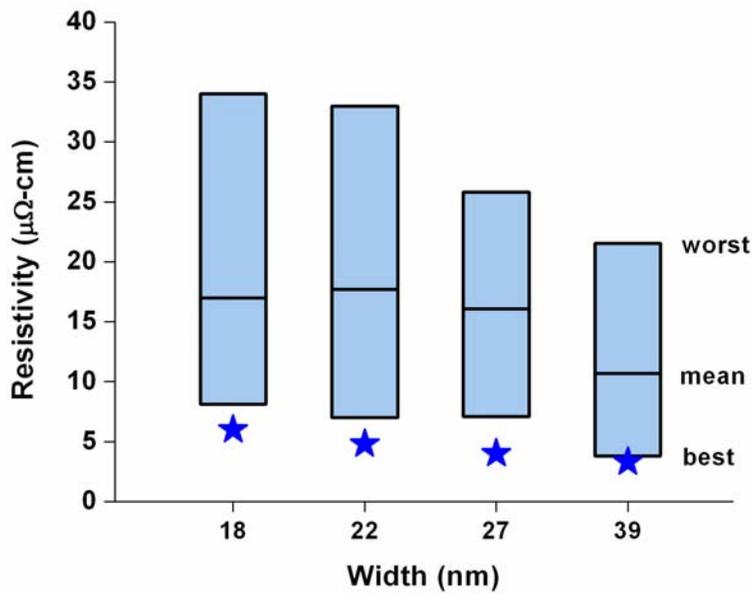

Figure 3: Performance of individual GNRs of various line-widths. The best, worst and mean values are shown for each line-width; there are ten data points at each width corresponding to ten GNRs between each electrode pair. The equivalent Cu resistivity is shown as a star.